\documentclass{emulateapj}

\usepackage{amsmath}
\usepackage{amssymb} 
\usepackage{eufrak}
\usepackage{graphicx}
\usepackage[colorlinks=true, linkcolor=blue, citecolor=blue, urlcolor=blue]{hyperref}
\usepackage{natbib}
\bibpunct{(}{)}{;}{a}{}{,}

\pdfoutput=1

\newcommand{\bfx}{{\boldsymbol{x}}}
\newcommand{\txd}{{\text{d}}}

\begin{document}

\title{Efficient 3D NLTE dust radiative transfer with SKIRT}

\author{\mbox{Maarten Baes}, \mbox{Joris Verstappen}, \mbox{Ilse De Looze}, \mbox{Jacopo Fritz},
  \mbox{Waad Saftly}, \mbox{Edgardo Vidal P\'erez}}
\affil{Sterrenkundig Observatorium, Universiteit Gent, Krijgslaan
  281-S9, B-9000 Gent, Belgium}
\email{maarten.baes@ugent.be} 
\and
\author{Marko Stalevski}
\affil{Astronomical Observatory Belgrade, Volgina 7, P.O.Box 74 11060
  Belgrade, Serbia}
\affil{Sterrenkundig Observatorium, Universiteit Gent, Krijgslaan
 281-S9, B-9000 Gent, Belgium}
\affil{Isaac Newton Institute, Yugoslavia branch}
\and
\author{Sander Valcke}
\affil{Applied Maths NV, Keistraat 120, 9830 Sint-Martens-Latem,
  Belgium}
\affil{Sterrenkundig Observatorium, Universiteit Gent, Krijgslaan
 281-S9, B-9000 Gent, Belgium}

\begin{abstract}
  We present an updated version of SKIRT, a 3D Monte Carlo
  radiative transfer code developed to simulate dusty galaxies. The
  main novel characteristics of the SKIRT code are the use of a
  stellar foam to generate random positions, an efficient combination
  of eternal forced scattering and continuous absorption, and a new
  library approach that links the radiative transfer code to the
  DustEM dust emission library. This approach enables a fast, accurate
  and self-consistent calculation of the dust emission of arbitrary
  mixtures of transiently heated dust grains and polycyclic aromatic
  hydrocarbons, even for full 3D models containing millions of dust
  cells. We have demonstrated the accuracy of the SKIRT code
  through a set of simulations based on the edge-on spiral galaxy
  UGC\,4754. The models we ran were gradually refined from a smooth,
  2D, LTE model to a fully 3D model that includes NLTE dust emission
  and a clumpy structure of the dusty ISM. We find that clumpy models
  absorb UV and optical radiation less efficiently than smooth models
  with the same amount of dust, and that the dust in clumpy models is
  on average both cooler and less luminous. Our simulations
  demonstrate that, given the appropriate use of optimization
  techniques, it is possible to efficiently and accurately run Monte
  Carlo radiative transfer simulations of arbitrary 3D structures of
  several million dust cells, including a full calculation of the NLTE
  emission by arbitrary dust mixtures.
\end{abstract}


\section{Introduction}

The effects of absorption and scattering by interstellar dust grains
on the structural parameters of galaxies has been a long-standing and
controversial issue. The only way to tackle this problem is to
properly solve the continuum radiative transfer equation, taking into
account realistic geometries and the physical processes of absorption
and multiple anisotropic scattering. Over the years, many different
and complementary approaches have been developed to tackle the
continuum radiative transfer problem in simple geometries such as
spherical or plane-parallel symmetry. While one-dimensional radiative
transfer calculations have been crucial to isolate and demonstrate the
often counter-intuitive aspects of important parameters such as
star-dust geometry, dust scattering properties and clumping
\citep{1988ApJ...333..673B, 1989MNRAS.239..939D, 1992ApJ...393..611W,
  1995MNRAS.277.1279D, 2001MNRAS.326..733B, 2005MNRAS.359..171I}, we
need more sophisticated radiative transfer models to model complicated
systems such as disc galaxies in detail. Thanks to new techniques and
ever increasing computing power, the construction of 2D and 3D
realistic radiative transfer models is now possible
\citep[e.g.][]{1999A&A...344..868X, 2000A&A...362..138P,
  2001ApJ...551..269G, 2003A&A...401..405S, 2005BaltA..14..543S,
  2006A&A...459..797P, 2006MNRAS.372....2J, 2008A&A...490..461B}.

A complementary and powerful way to study the content of galaxies is
to use the direct emission of dust at long wavelengths. Large dust
grains will typically reach a state of local thermal equilibrium (LTE)
in the local interstellar radiation field (ISRF) and re-radiate the
absorbed UV/optical radiation at far-infrared (FIR) and submm
wavelengths. Thanks to the spectacular advances in instrumentation in
the FIR/submm wavelength region, we have seen a significant
improvement in the amount of FIR/submm data on both nearby and distant
galaxies. In particular, the launch of the Herschel Space Observatory
with the sensitive PACS and SPIRE instruments has enabled both the
detailed study of nearby, resolved galaxies
\citep{2010A&A...518L..65B, 2010A&A...518L..55G, 2010A&A...518L..72P,
  2010A&A...518L..66R, 2010A&A...518L..51S} and the detection of
thousands of distant galaxies \citep{2010A&A...518L...8C,
  2010A&A...518L..21O, 2011MNRAS.tmp..955R}. Whereas large grains
typically emit as a modified blackbody at an equilibrium temperature
of 15-30~K and hence dominate the FIR/submm emission of galaxies,
small grains and PAH molecules are transiently heated by the
absorption of single UV photons to much higher temperatures. The NLTE
emission from very small grains and PAHs dominates the emission of
galaxies at mid-infrared wavelengths. The ISO and particularly the
Spitzer mission have been instrumental in uncovering the mid-infrared
emission of nearby galaxies \citep{2000ApJ...532L..21H,
  2007ApJ...656..770S, 2007ApJ...663..866D}.

Different approaches have been developed to calculate the NLTE
emission spectrum due to very small grains and PAHs
\citep{1986ApJ...302..363D, 1986A&A...160..295D, 1989ApJ...345..230G,
  1992A&A...266..501S, 2001ApJ...551..807D, 2011A&A...525A.103C}, but
the integration of NLTE emission into radiative transfer codes has
proven to be a challenging task. The main reason is that the
computational effort necessary to calculate the temperature
distribution of the different dust grains is substantial. In the
general case, the calculation of the dust emissivity in a single dust
cell requires the solution of a large matrix equation for each single
dust population, with the size of the matrix determined by the number
of temperature or enthalpy bins. In the so-called thermal continuous
cooling approximation \citep{2001ApJ...551..807D}, this matrix
equation can be solved recursively, but still the calculation of the
emission spectrum remains a significant computational
challenge. Indeed, since the temperature distribution of dust grains
depends strongly on both the size of the grains and the strength and
hardness of the ISRF, a large number of temperature or enthalpy bins
is necessary to sample the temperature distribution
correctly. Moreover, because of this strong dependence on grain size
and ISRF, the choice of the temperature bins is hard to fix a priori
and an iterative procedure is to be preferred.

In spite of the high numerical cost, NLTE dust emission has been built
into several radiative transfer codes, using various approximations
and/or assumptions. The most simple approach is the one followed by
e.g.\ \citet{2008ApJ...688.1118W} and \citet{2010MNRAS.403...17J}, who
use a set of predefined NLTE dust emissivities with the simplifying
assumption that the emissivity is a function only of strength and not
of the spectral shape of the exciting ISRF. A pioneering code in which
NLTE dust emission was included in a self-consistent way was the 2D
ray-tracing code by \citet{1992A&A...266..501S}. The number of
temperature distribution calculations are minimized by the assumptions
that grains with a size larger than about 80~\AA\ are in thermal
equilibrium, and by the use of a pre-fixed time dependence of the
cooling of PAH grains. A similar approach was adopted in the 3D Monte
Carlo radiative transfer code DIRTY \citep{2001ApJ...551..269G,
  2001ApJ...551..277M}. The TRADING code by
\citet{2008A&A...490..461B} uses a different approach: this code uses
a fixed (and limited) grid of temperature bins for all ISRFs and grain
sizes, which allows to precompute and tabulate a significant fraction
of the quantities necessary for the calculation of the temperature
distribution. Yet a different approach is the work by
\citet{2003A&A...397..201J}: driven by the observation that the
spectrum of the local ISRF is very similar in many places in a dusty
medium, they considered the idea of a dynamic library of dust emission
spectra. The idea is that the intensity of the ISRF at a very limited
number of reference wavelengths (they typically used only two)
suffices to make a reliable estimate of the total ISRF and hence of
the dust emission spectrum.

In this paper we present an updated version of the SKIRT Monte
Carlo radiative transfer code.  This code, of which the name is an
acronym to Stellar Kinematics Including Radiative Transfer, was
initially developed to study the effect of dust absorption and
scattering on the observed kinematics of dusty galaxies
\citep{2001ApJ...563L..19B, 2002MNRAS.335..441B,
  2003MNRAS.343.1081B}. In a second stage, the SKIRT code was
extended with a module to self-consistently calculate the dust
emission spectrum under the assumption of local thermal equilibrium
\citep{2005AIPC..761...27B, 2005NewA...10..523B}. This LTE version of
SKIRT has been used to model the dust extinction and emission
of various types of galaxies \citep{2010A&A...518L..45G,
  2010A&A...518L..39B, 2010A&A...518L..54D}, as well as circumstellar
discs \citep{2007BaltA..16..101V} and clumpy tori around active
galactic nuclei \citep{Stalevski2011, Popovic2011}. In this present
paper we present a strongly extended version of the SKIRT code
that can perform efficient 3D radiative transfer calculations
including a self-consistent calculation of the dust temperature
distribution and the associated FIR/submm emission with a full
incorporation of the emission of transiently heated grains and PAH
molecules.

In Section~{\ref{SKIRT-general.sec}} we present the general
characteristics of the SKIRT code, whereas we highlight a
number of particular aspects in Section~{\ref{particular.sec}} and
some implementation details in Section~{\ref{implementation.sec}}. In
Section~{\ref{applications.sec}} we describe a number of tests and
applications, and Section~{\ref{conclusions.sec}} we present our
conclusions.

\section{The SKIRT Monte Carlo radiative transfer code}
\label{SKIRT-general.sec}

\subsection{Monte Carlo radiative transfer}

SKIRT is a 3D continuum radiative transfer code based on the
Monte Carlo algorithm. The key principle in Monte Carlo radiative
transfer simulations is that the radiation field is treated as a flow
of a finite number of photon packages. A simulation consists of
consecutively following the individual path of each single photon
package through the dusty medium. The journey or lifetime of a single
photon package can be thought of as a loop: at each moment in the
simulation, a photon package is characterized by a number of
properties, which are generally updated when the photon package moves
to a different stage on its trajectory. The trajectory of the photon
package is governed by various events such as emission, absorption and
scattering events. Each of these events is determined statistically by
random numbers, generated from the appropriate probability
distribution $p(x)\, \txd x$. Typically, a photon is emitted by a
star, undergoes a number of scattering events and its journey
ultimately ends when it is either absorbed by the dust or it leaves
the system. A Monte Carlo simulation repeats this same loop for every
single one of the photon packages and analyzes the results afterwards.

The mathematical details and practical implementation of Monte Carlo
radiative transfer have both been described in detail by various
authors \citep[e.g.][]{CashEv, 1970A&A.....9...53M,
  1977ApJS...35....1W, 1994A&A...284..187F, 1996ApJ...465..127B,
  2001ApJ...551..269G, 2003A&A...399..703N, 2003CoPhC.150...99W,
  2003A&A...407..941S, 2005A&A...440..531J, 2006MNRAS.372....2J,
  2008A&A...490..461B} and will not be repeated here in full
detail. Our overall approach is comparable to the DIRTY
\citep{2001ApJ...551..269G, 2001ApJ...551..277M} and TRADING
\citep{2008A&A...490..461B} radiative transfer codes and we refer the
interested reader to these papers for more details. We will only give
a compact description of the general characteristics of the
SKIRT Monte Carlo code and not describe all the
details. Instead, we will focus our attention to those aspects of the
SKIRT code that are novel and/or different compared to the
other codes.

\subsection{General overview of a SKIRT simulation}

Each SKIRT simulation consists of four phases: the
initialization phase, the stellar emission phase, the dust emission
phase, and the clean-up phase.

\subsubsection{The initialization phase}

The initialization phase consists of adopting the correct unit system,
setting up the random number generator, computing the optical
properties of the various dust populations, constructing the dust
grid, setting up the stellar geometry and setting up the instruments
of the various observers. Once this initialization is finished, the
actual simulation can start.

\subsubsection{The stellar emission phase}

In the stellar emission phase, we consider the transfer of the primary
source of radiation (usually stellar sources, but it can also include
an accretion disc or nebular line emission) through the dusty
medium. The stellar emission phase consists of a set of parallel
loops, each of them corresponding to a single wavelength. At every
single wavelength, the total luminosity of the stellar system is
divided into a very large number (typically $10^5$ to $10^7$) of
monochromatic photon packages, which are launched consecutively
through the dusty medium in random propagation directions.

Once a photon package is launched into the dusty medium (either after
an emission event or following a scattering event), it can be absorbed
by a dust grain, it can be scattered by a dust grain, or it can travel
through the system without any interaction. In a naive Monte Carlo
routine, these three possibilities are possible and it is randomly
determined which of the three will happen. This is generally an
inefficient procedure, though, which leads to poor signal-to-noise
both in the absorption rates in the different cells and in the
scattered light images. To overcome these problems, we have set up a
combination of continuous absorption and eternal forced scattering
(see Section~{\ref{forced.sec}} for details). The result is that,
contrary to most Monte Carlo codes where the life cycle of a photon
package ends when it either leaves the system or is absorbed, the
photon packages in SKIRT can never leave the system. The life
cycle of a photon package ends when the package contains virtually no
more luminosity (typically we use the criterion that it must have lost
99.99\% of its original luminosity). Whenever this happens, a
different stellar photon package is launched until also this one
finishes its life cycle. This loop is repeated for all stellar photon
packages at a given wavelength, and subsequently for all wavelengths
(in a multi-core system, each core can handle the loop corresponding
to a different wavelength at the same time).

\subsubsection{The dust emission phase}

After the stellar emission phase, the code moves to the dust emission
phase. This phase is roughly similar to the stellar phase, except that
the sources that emit the radiation are now not the primary, stellar
sources but the dust cells. From the stellar emission phase we know
the total amount of absorbed radiation at each wavelength in each cell
of the dust domain. From this absorption rate we can calculate the
mean intensity of the ISRF in each cell, which allows the calculation
of the dust emissivity, depending on the physical processes the
SKIRT user is interested in (see
Section~{\ref{dustemissivity.sec}}).  With the sources (the dust
cells) and their emissivity determined, the simulation now enters a
loop that is very similar to the one in the stellar emission phase. At
each individual wavelength, a huge number of photon packages is
generated which are launched and followed consecutively through the
dusty medium. Care is taken that all regions of the dusty medium,
including those cells where only a small amount of luminosity has been
absorbed, are well-sampled.

The dust-emitted photon packages in turn increase the absorption rate
in the dust cells where they pass through. This results in an increase
of the mean intensity of the ISRF. The result is that at the end of
the dust emission phase, the absorption rates used to calculate the
dust emissivity in each cell do not correspond to the mean intensity
of the ISRF. This naturally leads to an iterative procedure, in which
the absorption rate, the mean intensity and the dust emissivity are
updated until convergence is achieved \citep{2001ApJ...551..277M,
  2008A&A...490..461B}. We hence repeat the dust emission phase of the
code several times. We typically require a 1\% level accuracy in the
dust bolometric absorption rate of each cell as a stopping criterion
for the iteration. It is typically reached in only a few iterations;
for all the simulations we have done so far, less then five iterations
have been necessary.

\subsubsection{The clean-up phase}

The last phase of the Monte Carlo simulation starts when the last of
the photon packages emitted by the dust component has lost 99.99\% of
its initial luminosity. It simply consists of calibrating and reading
out the instruments (all output is written to FITS files) and other
useful information, such as 3D absorption rate maps and dust
temperature distributions.

\section{Particular aspects of the SKIRT code}
\label{particular.sec}

\subsection{Setup of the dust grid}
\label{dustgrid.sec}

A critical aspect in Monte Carlo radiative transfer simulations is the
choice of the dust grid. The dust grid consists of tiny cells, each of
which have a number of characteristics that fully describe the
physical properties of the dust at the location of the cell. The
choice of the grid has a significant impact on both the run time and
the memory requirement of the simulation. Indeed, each photon package
typically requires several integrations through the dust (i.e. the
determination of the optical depth along the path and the conversion
of a given optical depth to a physical path length), and the
calculation time of a single optical depth typically scales with the
number of grid cells crossed. Different kinds of dust grids can be
applied in the SKIRT code. The most general grid is a 3D
cartesian grid in which each dust cell is a rectangular cuboid. For
simulations with a spherical or axial symmetry, we also have 1D
spherical and 2D cylindrical grids (the elementary dust cells being
shells or tori respectively). The distribution of the grid points (in
1D spherical, 2D cylindrical or 3D cartesian grids) can be chosen
arbitrarily; linear, logarithmic, or power-law cell distributions have
been pre-programmed, but any user-supplied grid cell distribution is
possible.

The main goal of the dust grid is to discretize the dust density. We
assume that the density of each dust component is uniform within each
individual cell. In principle, the density does not need to be
constant within each dust cell \citep[see
e.g.][]{2003A&A...399..703N}. In the first versions of SKIRT,
we have experimented with a more sophisticated kind of dust grid,
where the density of the dust within each cell is not uniform but
determined by trilinear interpolation of the values of the density on
the eight border points of the cell (in case of a cartesian grid with
cubic cells). In this case, the computation of optical depths in the
dusty medium take more computation time, but the accuracy is increased
such that a grid with less cells and less photon packages are needed
per simulation. For models in which only absorption and scattering are
taken into account, we found that this kind of dust grid is
computationally more efficient than a dust grid with uniform density,
in particular when the system harbors a large dynamical range of dust
densities \citep{2003MNRAS.343.1081B}. However, for radiative transfer
simulations in which the thermal emission of the dust is taken into
account, each dust cell needs to contain an absorption rate counter,
which collects the absorbed luminosity at every wavelength. The size
of the dust cells is hence the typical resolution of the simulation,
and the advantage of the interpolated grid (where the dust grid can be
coarser because the density is resolved within each cell) largely
disappears. SKIRT therefore only uses dust grids with a uniform
density in each cell.

\subsection{Sampling the stellar density}
\label{stellarfoam.sec}

The first step in the life cycle of each stellar photon is the random
generation of the location where it is emitted. This means that we
have to generate random positions from the three-dimensional
probability distribution
\begin{equation}
  p(\bfx)\,\txd\bfx
  =
  \frac{\nu_\lambda(\bfx)\,\txd\bfx}{\int \nu_\lambda(\bfx')\,\txd\bfx'}
\end{equation}
where $\nu_\lambda(\bfx)$ is the luminosity density at the wavelength
$\lambda$ of the photon package. As SKIRT is primarily focused
towards modeling galaxies, we have done efforts to optimize the
generation of random positions from arbitrary 3D probability
functions. The SKIRT code contains a library with common
geometries for which the generation of a random position vector can be
performed analytically. These include spherical
\citet{1911MNRAS..71..460P}, \citet{1983MNRAS.202..995J} or
\citet{1990ApJ...356..359H} models, or axisymmetric power-law,
exponential, sech or isothermal disc models. For other frequently used
luminosity density profiles, e.g.\ flattened
\citet{1948AnAp...11..247D} or \citet{1968adga.book.....S} models, or
more general density profiles that cannot be described by an
analytical function, such a direct analytical inversion is not
possible. Two complementary approaches have been included in the
SKIRT code to deal with generating random positions from such
density profiles.

\subsubsection{Multi-gaussian expansion technique}

\begin{figure}
\centering
\includegraphics[width=0.48\textwidth]{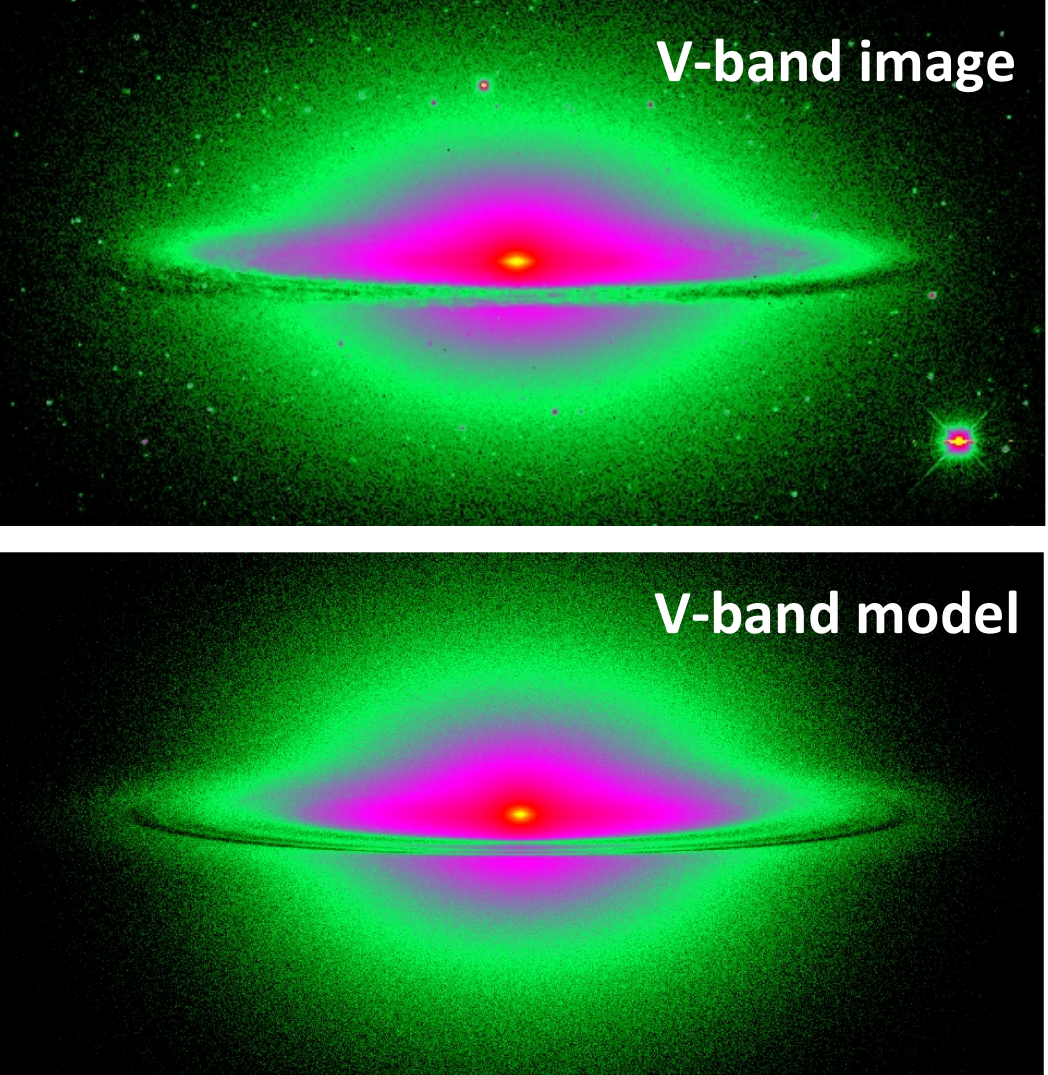}
\caption{Observed (top) and simulated (bottom) V-band images of the
  Sombrero Galaxy (M104). The observed image is taken from the Spitzer
  Infrared Nearby Galaxies Survey
  \citep[SINGS,][]{2003PASP..115..928K} ancillary data website. In the
  SKIRT simulation, the stellar density profile is based on a
  multi-gaussian expansion (MGE) with 15 components, as presented by
  \citet{1995A&A...303..673E}. For a full-scale panchromatic radiative
  transfer modeling of stars and dust in M104, see
  \citet{DeLooze2011}.}
\label{Sombrero.pdf}
\end{figure}

The first technique is to expand the density profile into a set of
subcomponents. SKIRT contains a routine to perform a
multi-gaussian expansion (MGE) of surface brightness distributions. An
MGE expansion basically expands any surface brightness distribution as
a finite sum of two-dimensional gaussian components
\citep{1992A&A...253..366M, 1994A&A...285..723E,
  2002MNRAS.333..400C}. The MGE method has proven to be a very
powerful tool for image analysis: even with a relatively modest set of
gaussian components, $N\sim10$, even complex geometries can be
reproduced accurately \citep{1994A&A...285..739E, 1999MNRAS.303..495E,
  2002MNRAS.333..400C, 2006MNRAS.366.1126C, 2008MNRAS.385..647V}. One
of the reasons why a MGE expansion is very useful for SKIRT is
that this approach enables a straightforward determination of the 3D
spatial distribution: if the Euler angles of the line of sight are
known, the de-projection of a 2D gaussian distribution on the plane of
the sky is a 3D gaussian distribution and the conversion formulae are
completely analytical \citep{1992A&A...253..366M}. Generating random
positions from a sum of 3D gaussian probability distributions is
straightforward. An example of this approach can be seen in
Figure~{\ref{Sombrero.pdf}}, where we present a radiative transfer
model for the Sombrero Galaxy based on the MGE expansion of its
surface brightness distribution presented by
\citet{1995A&A...303..673E}.

\subsubsection{The stellar foam}

The second approach consists of sampling random positions directly
from the stellar density distribution using the so-called stellar
foam.  The stellar foam is a SKIRT structure based on the Foam library
developed by \citet{2003CoPhC.152...55J} and
\citet{2007CoPhC.177..441J}. Foam is a self-adapting cellular Monte
Carlo tool aimed at Monte Carlo integration of multi-dimensional
functions, including integrands with an arbitrary pattern of
singularities. It achieves a high efficiency thanks to an intelligent
division of the integration space into small simplical or
hyper-rectangular form, which are created in a self-adaptive way by
binary splitting. It has originally been developed for use in
high-energy physics \citep[e.g.][]{2006CoPhC.175..511J,
  2010PAN....73.1761A, 2011EPJC...71.1574H}, but it can also be
adopted as a general-purpose Monte Carlo event generator.

We use an adapted version of the Foam library for the generation of
random positions from an arbitrary probability density
$p(\bfx)\,\txd\bfx$. One problem is that the probability density
$p(\bfx)\,\txd\bfx$ from which we want to generate random positions is
typically defined on the entire 3D space, whereas Foam requires a
probability density on the $N$-dimensional unit hypercube
$[0,1]^N$. We achieve this through a coordinate transformation from
the usual coordinates $\bfx=(x,y,z)$ to new coordinates $\bar{\bfx} =
(\bar{x}, \bar{y}, \bar{z})$ such that we map each infinite interval
$[-\infty,\infty]$ onto the unit interval $[0,1]$.  The probability
density $p(\bfx)\,\txd\bfx$, normalized such that
\begin{equation}
  \int_{-\infty}^\infty \txd x
  \int_{-\infty}^\infty \txd y
  \int_{-\infty}^\infty p(x,y,z)\, \txd z 
  = 
  1
\end{equation}
is transformed to a new probability density 
\begin{equation}
\label{barp}
  \bar{p}(\bar\bfx)\,\txd\bar\bfx 
  =
  p\Bigl( x(\bar{x}), y(\bar{y}), z(\bar{z}) \Bigr)\,
  \left|
  \frac{{\txd}x}{{\txd}\bar{x}}\,
  \frac{{\txd}y}{{\txd}\bar{y}}\,
  \frac{{\txd}z}{{\txd}\bar{z}}
  \right|\,
 \txd\bar\bfx
\end{equation} 
This new density will be defined and normalized to one on the unit
cube,
\begin{equation} 
  \int_0^1 \txd\bar{x} 
  \int_0^1 \txd\bar{y} 
  \int_0^1 \bar{p}(\bar{x},\bar{y},\bar{z})\, {\txd}\bar{z} 
  = 
  1 
\end{equation}
There are many possible transformations we can apply to achieve this. We
have chosen the transformation
\begin{subequations}
\label{barxyz}
\begin{gather} 
  \bar{x} = \frac{1}{\pi} \arctan\left(\frac{a}{x}\right) 
  \\ 
  \bar{y} = \frac{1}{\pi} \arctan\left(\frac{b}{y}\right) 
  \\ 
  \bar{z} = \frac{1}{\pi} \arctan\left(\frac{c}{z}\right) 
\end{gather} 
\end{subequations}
with $a$, $b$ and $c$ three scale parameters, which we can adapt for
the specific probability density we are considering. The inverse
transformation is 
\begin{subequations}
\label{xyz}
\begin{gather} 
  x = \frac{a}{\tan(\bar{x}\pi)} 
  \\ 
  y = \frac{b}{\tan(\bar{y}\pi)} 
  \\
  z = \frac{c}{\tan(\bar{z}\pi)}
\end{gather} 
\end{subequations}
and the Jacobian reads 
\begin{equation}
\label{jacobian} 
  \left|\frac{\partial(x,y,z)}{\partial(\bar{x},\bar{y},\bar{z})} \right|
  = 
  \left| 
    \frac{{\txd}x}{{\txd}\bar{x}}\,
    \frac{{\txd}y}{{\txd}\bar{y}}\, 
    \frac{{\txd}z}{{\txd}\bar{z}}
  \right| 
  = 
  \frac{abc\,\pi^3}{\sin^2(\bar{x}\pi) 
    \sin^2(\bar{y}\pi) \sin^2(\bar{z}\pi)} 
\end{equation}
Summarizing, if we want to generate random positions $\bfx$ from an
arbitrary probability density $p(\bfx)\,\txd\bfx$, we first determine
representative scale lengths $a$, $b$ and $c$ along the three
dimensions, and subsequently calculate the corresponding probability
density $\bar{p}(\bar\bfx)\, \txd\bar\bfx$ using
equations~(\ref{barp}), (\ref{xyz}) and (\ref{jacobian}). The foam
generator is applied to this new probability function to generate the
random points $\bar\bfx$, which are converted to the desired positions
$\bfx$ through the formulae~(\ref{xyz}). The construction of the
stellar foam takes only a few seconds in two dimensions up to about
one minute in three dimensions.

\subsection{Eternal forced scattering and continuous absorption}
\label{forced.sec}

Once a photon package has been generated at a random location (sampled
from the stellar density) and it has been given a random propagation
direction (sampled randomly from the unit sphere), it is ready to
start its journey through the dusty medium. It has three possible
fates: it can be absorbed by a dust grain at a certain position along
its path, it can be scattered by a dust grain at a certain position on
its path, or it can travel along its path through the system without
any interaction. The probability for each of these three options are
determined by the scattering albedo $\varpi_\lambda$ and the optical
depth $\tau_{\lambda,{\text{path}}}$ along the path, defined as
\begin{equation}
  \tau_{\lambda,\text{path}} 
  = 
  \int_0^\infty \kappa_\lambda\,\rho(s)\,{\text{d}}s
\end{equation}
where $\kappa_\lambda$ is the extinction coefficient, $\rho$ is the
dust density and the integral covers the entire path of the photon
package through the dusty medium. The most straightforward way to
model these different physical processes in a Monte Carlo radiative
transfer code is to generate randomly which of these three processes
will take place. In case the photon package leaves the system, its
lifetime is terminated and a different package is launched. In the
case of a scattering event, the position of the scattering event is
determined by choosing a random path length from the appropriate
probability distribution and a scattering event is simulated. Finally,
in case it is absorbed, the position of the absorption is determined
in a similar way, the luminosity of the photon package is stored in a
local absorption rate counter attached to the dust cell where the
absorption event took place, and the photon package's life is
terminated. These local absorption rates are used in a later stage of
the simulation to estimate the local mean intensity of the ISRF,
necessary to calculate the thermal dust emission.

This traditional method has two significant drawbacks: along paths
where the optical depth is modest, many photon packages will escape
from the system without interactions, which will result in bad
statistics of the scattered intensity and the absorbed
luminosity. Even if the photon packages do interact, most interactions
will take place on those sections of the path where the density is
largest. Many absorption events are necessary in each cell to
guarantee a high-quality estimate of the local absorption rate and the
corresponding mean intensity. In dust cells with a low density (such
that only few absorptions take place) and at wavelengths where the
absorption rate is low, this usually requires large numbers of photon
packages. 

These two problems can be minimized using the efficient combination of
two clever Monte Carlo techniques: forced scattering and continuous
absorption. Continuous absorption (or Lucy estimation) is a technique
to estimate the mean intensity of the ISRF throughout the dusty medium
\citep{1999A&A...344..282L, 2003A&A...399..703N}. The continuous
absorption technique is designed to solve the problem of poor
statistics in the absorption rate in low-density regions by spreading
the absorption over all cells along the photon package's path instead
of concentrating it in one single cell. A different but equivalent way
to see it is that the mean intensity in each cell is estimated using
the sum of the path lengths covered by all photon packages that
traverse that particular cell. Forced scattering is an old technique
that was already implemented in the first Monte Carlo radiative
transfer codes \citep{CashEv, 1970A&A.....9...53M,
  1977ApJS...35....1W}. When applying forced scattering, photon
packages are forced to interact with a dust grain before they leave
the system. This incorrect behavior is corrected for by decreasing
the luminosity of the photon package.

In most radiative transfer codes, forced scattering is considered only
after an emission event and subsequent scattering events are
unforced. In the SKIRT code we always consider forced scattering, such
that we have eternal forced scattering. The combination of eternal
forced scattering and continuous absorption results in a very
efficient Monte Carlo routine. Instead of determining randomly whether
a photon package with luminosity $L_\lambda$ will escape, will be
absorbed or will be scattered, we split it into $n+2$ child photon
packages (with $n$ the number of dust cells along the path): one child
photon package that will leave the system without interaction, one
child photon package that will be scattered by a dust grain somewhere
along the path, and $n$ children that will be absorbed, one in each of
the $n$ cells along the path. The luminosity of each of these children
is easy to calculate: we find
\begin{gather}
  L_{\lambda,\text{abs}_1}
  =
  L_\lambda\,(1-\varpi_\lambda)
  \left(1-{\text{e}}^{-\tau_{\lambda,1}}\right)
 \\
  L_{\lambda,\text{abs}_2}
  =
  L_\lambda\, (1-\varpi_\lambda)
  \left({\text{e}}^{-\tau_{\lambda,1}}-{\text{e}}^{-\tau_{\lambda,2}}\right)
 \\
  \vdots
  \nonumber \\
  L_{\lambda,\text{abs}_{n-1}}
  =
  L_\lambda\,(1-\varpi_\lambda)
  \left({\text{e}}^{-\tau_{\lambda,n-2}}-{\text{e}}^{-\tau_{\lambda,n-1}}\right)
  \\
  L_{\lambda,\text{abs}_n}
  =
  L_\lambda\,(1-\varpi_\lambda)
  \left({\text{e}}^{-\tau_{\lambda,n-1}}-{\text{e}}^{-\tau_{\lambda,\text{path}}}\right)
  \\
  L_{\lambda,\text{sca}}
  =
  L_\lambda\,(1-\varpi_\lambda)\left(1-{\text{e}}^{-\tau_{\lambda,\text{path}}}\right)
  \\
  L_{\lambda,\text{esc}}
  =
  L_\lambda\,{\text{e}}^{-\tau_{\lambda,\text{path}}}
\end{gather}
with $\tau_{\lambda,j}$ the optical depth to the point on the path
where the photon package would leave the $j$'th dust cell. Obviously,
we have
\begin{equation}
  L_{\lambda,\text{esc}}
 +
  L_{\lambda,\text{sca}}
 +
  \sum_{j=1}^m L_{\lambda,\text{abs}_j}
 =
 1
\end{equation}
In our Monte Carlo simulation, we now consider each of these $n+2$
child photon packages. Each one of the $n$ children with luminosity
$L_{\lambda,\text{abs}_j}$ is absorbed in the $j$'th cell along the
path, which means that its luminosity is added to the absorption rate
counter attached to this dust cell. The child photon package with
luminosity $L_{\lambda,\text{esc}}$ escapes from the system, which
implies that we don't have to take this one into account
anymore. Finally, the child photon package with luminosity
$L_{\lambda,\text{sca}}$ is scattered somewhere along the path. This
is basically the only photon package that we need to follow up. We
still have to randomly determine the location of the scattering event
along the path. This is achieved by selecting a random optical depth
$\tau_\lambda$ from an exponential distribution over the finite range
$0<\tau_\lambda<\tau_{\lambda,\text{path}}$, i.e.\
\begin{equation}
  p(\tau_\lambda)\,{\text{d}}\tau_\lambda
  =
  \frac{{\text{e}}^{-\tau_\lambda}\,{\text{d}}\tau_\lambda}
  {1-\text{e}^{-\tau_{\lambda,\text{path}}}}
\end{equation}
and translate this randomly generated $\tau_\lambda$ to a physical
path length by solving the equation
\begin{equation}
  \tau_\lambda 
  = 
  \int_0^s \kappa_\lambda\,\rho(s')\,{\text{d}}s'
\end{equation}
for $s$. Once this path length has been determined, we can determine
the position of the scattering event. If we then also determine a new
propagation direction, determined randomly by generating a random
direction from the scattering phase function, we are back at the
starting point. This child now becomes the parent, it can be split
into children and we can repeat the same loop all over again.

Summarizing, the net result of the combination of continuous
absorption and eternal forced scattering is that after each
emission/scattering event, we distribute a fraction of the photon
package's luminosity among all the cells along the path, and we
continue the Monte Carlo loop with a less luminous photon package that
is always scattered at some point along the path. Hence, contrary to
most Monte Carlo codes where the life cycle of a photon package ends
when it either leaves the system or is absorbed, the photon packages
in SKIRT can never leave the system. The life cycle of a photon
package ends when the package contains virtually no more
luminosity. Typically we use the criterion that it must have lost
99.99\% of its original luminosity, which immediately is the minimum
level of absolute energy conservation of the simulation.

\subsection{Peeling off and smart detectors}
\label{smartdetector.sec}

The goal of a Monte Carlo radiative transfer simulation is to simulate
observable properties of a dusty system, i.e.\ images and spectral
energy distributions. SKIRT uses the technique of peel-off
photon packages to create an arbitrary number of images/SEDs at
different observing positions. Peeling off is a Monte Carlo technique
designed to create high signal-to-noise images (in particular
scattered light images), adopted for the first time by
\citet{1984ApJ...278..186Y} and included in almost all
state-of-the-art Monte Carlo radiative transfer codes. After every
emission or scattering event, the code calculates which fraction of
the luminosity contained in the photon package would arrive at the
observers' locations and at which point on the plane of the sky, if
the photon package would be emitted or scattered in the appropriate
propagation direction. Repeating this for every photon package at
every emission or scattering event implies that the maximum available
information is obtained for a fixed set of photon packages and hence
strongly increases the signal-to-noise compared to the more simple
Monte Carlo codes where only photon packages that leave the system are
recorded.

Each SKIRT detector is basically an integral field detector,
i.e.\ a data cube with two spatial and one wavelength dimension. In
most Monte Carlo radiative transfer codes the simulated detectors are
natural, idealized representations of actual CCD detectors (or a
series of them at each wavelength).  They basically consist of a two-
or three-dimensional array of pixels, which act as a reservoir for the
incoming photon packages. When a photon package leaves the system and
arrives at the location of the observer, the correct pixel is
determined and the luminosity of the photon package is added to the
luminosity at that pixel. At the end of the simulation, the detector
is read out pixel by pixel and the surface brightness distribution is
constructed. While this approach seems the most natural way to
simulate the detection of photon packages in a Monte Carlo simulation,
it might not be the most efficient. We must be aware that, although we
are simulating a real detection as closely as possible, we have more
information at our disposal than real observers. The maximum
information that a real observer can obtain (in the theoretical limit
of perfect noise-free observations and instruments) when imaging with
a CCD detector is the number of photon packages that arrive in each of
his pixels. As numerical simulators, we have at our disposal the full
information on the precise location of the impact of each photon
package on the detector. In order to use this information, we have
considered the concept of smart detectors, which take full advantage
of the exact location of the impact of the incoming photon packages
\citep{2008MNRAS.391..617B}. The principle of these smart detectors is
based on the estimate of the density distribution in smoothed particle
hydrodynamics simulations. While preserving the same effective
resolution and having virtually no computational overhead, smart
detectors realize a noise reduction of about 10 percent.

\subsection{The dust emissivity}
\label{dustemissivity.sec}

A crucial aspect of the SKIRT code is the calculation of the dust
emissivity.  From the stellar emission phase we know the total amount
of absorbed radiation $L_{\lambda,m}^{\text{abs}}$ at each wavelength
in each cell of the dust domain. From this absorption rate we can
calculate the mean intensity of the ISRF $J_{\lambda,m}$ in each cell
using
\begin{equation}
  J_{\lambda,m}
  =
  \frac{L_{\lambda,m}^{\text{abs}}}
  {4\pi\,\kappa_\lambda^{\text{abs}}\rho_m\,V_m}
\end{equation}
where $\kappa_\lambda^{\text{abs}}$ is the dust absorption
coefficient, $\rho_m$ is the dust density and $V_m$ is the volume in
cell number $m$ respectively. Knowledge of the mean intensity and the
dust properties in each cell allows the dust emissivity
$j_{\lambda,m}^\txd$ to be determined. SKIRT contains three different
modules for the calculation of the dust emissivity, depending on the
physical processes that are taken into account.

\subsubsection{Three different options for the dust emissivity}

The simplest option is to consider the dust grains as a single species
that is in local thermal equilibrium (LTE) with the local ISRF. In
this case, the dust emits as a modified blackbody radiator,
\begin{equation}
   j_{\lambda,m}^\txd 
   =
   4\pi\,\rho_m\,V_m\,\kappa_\lambda^{\text{abs}}\,B_\lambda(T)
 \end{equation}
 where the dust equilibrium temperature $T_m$ of the $m$'th dust cell
 is determined by the condition of thermal and radiative equilibrium,
\begin{equation}
  \int_0^\infty\kappa_\lambda^{\text{abs}} J_{\lambda,m}\,\txd\lambda
  =
  \int_0^\infty\kappa_\lambda^{\text{abs}} B_\lambda(T_m)\,\txd\lambda
\end{equation}
The second, somewhat more realistic option is to still consider LTE
for the dust grains, but taking into account that each species and
size of dust grain reaches its own equilibrium temperature. The
SKIRT code allows to consider dust mixtures with an arbitrary
number of grain species and size distributions. The size distributions
of the various dust species are subdivided into different bins,
resulting in a dust mixture with $N_{\text{pop}}$ populations, each of
them corresponding to a dust species and a small size bin. Assuming
LTE for each individual population, the dust emissivity is given by a
sum of modified blackbodies, where the temperature of each population
is still determined by the condition of thermal and radiative
equilibrium,
\begin{equation}
  j_{\lambda,m}^\txd 
  =
  4\pi\,\rho_m\,V_m
  \sum_{i=1}^{N_{\text{pop}}}\kappa_{\lambda,i}^{\text{abs}}\,B_\lambda(T_{m,i})
\label{jd-LTE}
\end{equation}
with $\kappa_{\lambda,i}^{\text{abs}}$ is the contribution of the
$i$'th dust population to the total absorption coefficient
$\kappa_\lambda^{\text{abs}}$, and $T_{m,i}$ the equilibrium
temperature of the $i$'th population in cell number $m$, determined by
\begin{equation}
  \int_0^\infty\kappa_{\lambda,i}^{\text{abs}} J_{\lambda}\,\txd\lambda
  =
  \int_0^\infty\kappa_{\lambda,i}^{\text{abs}} B_\lambda(T_{m,i})\,\txd\lambda
\end{equation}
The third option, in fact the only realistic option to model the
spectral energy distribution of galaxies, is to consider NLTE dust
emission. In theory, the transit from LTE to NLTE dust emission is not
an enormous step. The main difference is that each dust population is
not characterized by a single equilibrium temperature, but by a
temperature distribution. Once the temperature distribution function
has been determined, the dust emissivity can easily be determined. As
argued in the Introduction, however, the practical inclusion of NLTE
dust emission in radiative transfer codes is a notoriously tough nut
to crack.

Rather than develop our own routines to calculate the NLTE emission
for transiently heated grains, we have opted to couple the
SKIRT code to the DustEM code
\citep{2011A&A...525A.103C}. DustEM is a publicly available,
state-of-the-art numerical tool designed to calculate the NLTE
emission and extinction of dust given its size distribution, optical
and thermal properties. The code builds on the work by
\citet{1986A&A...160..295D, 1990A&A...237..215D} and uses an adaptive
temperature grid on which the temperature distribution of the grains
is calculated iteratively. No LTE approximation is made, i.e.\ even
for large grains the temperature distribution is calculated
explicitly. One of the advantages why we have chosen to couple
SKIRT to the DustEM code is that the latter code has been
designed to deal with a variety of grain types, structures and size
distributions and that new dust physics (ionization of PAHs, polarized
emission, spinning dust emission, temperature-dependent dust
emissivity) can easily be included.  On the other hand, a consequence
of choosing for a very complete and accurate NLTE routine in which
basically no simplifications or assumptions have been made, is that
the computation load is substantial. For a typical dust model
consisting of three or four dust species each with their size
distribution, the calculation of the emissivity for a single ISRF
takes typically of the order of several seconds on a standard
desktop/laptop computer. While this is compatible with 1D or 2D
simulations with up to $10^4$ cells, this is excessively long for
general 3D simulations for which we have designed SKIRT.

\subsubsection{A library approach for NLTE dust emission}
 
To overcome this problem, we have adopted a strategy based on a
library of dust emissivity profiles, inspired by the work of
\citet{2003A&A...397..201J}. Their approach consists of three steps:
they first run an exploratory radiative transfer simulation on a grid
with a reduced number of grid cells, without taking dust re-emission
into account. This low-resolution simulation is used to determine the
range of ISRFs encountered in the simulation. The second step consists
of picking a small number $N_{\text{ref}}$ of reference wavelengths
$\lambda^{\text{ref}}_i$ (typically $N_{\text{ref}}=2$). The different
ISRFs found in step one are discretized onto logarithmic intervals at
each of the reference wavelengths, and at each bin in the
$N_{\text{ref}}$-dimensional parameter space, the full NLTE dust
emission spectrum is calculated and stored in a library. The final
step in the simulation consists of running a radiative transfer
simulation at the full resolution. The dust emissivity at any given
cell is found by looking at the ISRF at the $N_{\text{ref}}$ reference
wavelengths and interpolating the dust emissivities from the library.

While valuable and inspiring, we see two drawbacks in the method as
implemented by \citet{2003A&A...397..201J}. In panchromatic radiative
transfer simulations of galaxies, we typically solve the transfer
equation at many UV/optical wavelengths, and hence have the ISRF at
all these wavelengths at our disposal at every grid cell. It would be
a pity not to use this information to determine the dust emissivity
and only base our estimate on the value of the ISRF at a very small
number of reference wavelengths. In particular, the ISRF in Monte
Carlo simulations can be noisy in certain cells; when the dust
emissivity is determined based on the value of the ISRF at a small
number of reference wavelengths, this noise could lead to a
significant error. Using an estimate that exploits the available
information at all wavelengths can minimize this error.

The second drawback is that the library method of
\citet{2003A&A...397..201J} requires an exploratory, low-resolution
simulation in which the parameter space of ISRFs is explored and the
library of dust emissivities is built. This extra simulation not only
requires a computational overhead, it also creates the danger that it
does not cover the entire range of strengths and shapes of the
ISRF. For example, one can assume that the strongest ISRF in a
simulation is found in small dust cells very close to the heating
sources. In a low-resolution simulation, with larger dust cells, this
strong ISRF will be smoothed over the larger grid cells. Similarly,
the weakest ISRF (or equivalently, the coldest dust) in some
simulations could be found in the inner regions of dense cores, and
due to smoothing a low-resolution simulation might not reach these
weakest ISRF levels. The result is that the low-resolution grid will
not cover the full range of ISRFs encountered in the high-resolution
grid, and hence that the library of dust emissivities must be somehow
extended to incorporate this missing part in the parameter
space. \citet{2003A&A...397..201J} are aware of this inconvenience
(they discuss only the coldest spectrum as they concentrate on dark
clouds illuminated by an external radiation field). They argue that
this problem is not expected to be significant, and that it could be
relieved by using a low-resolution simulation with a slightly higher
density. Still, it is clear that the use of a low-resolution grid
leads to an additional overhead and complication and is a potential
source of error, and it would be better to avoid it.

To overcome both problems, we have taken a slightly different approach
to implement our dust emissivity library. The first step in our
library approach is to calculate a number of parameters that
characterize the ISRF in each dust cell after the stellar emission
phase. Instead of the value of $J_\lambda$ at a number of wavelengths,
we use parameters that use combined information at all available
wavelengths. From the various range of possibilities, we choose the
lowest-order moments of $\kappa_\lambda^{\text{abs}}J_\lambda$, the
product of the ISRF and the dust absorption coefficient. Instead of
the actual zeroth-order moment or normalization of this function, we
consider the equivalent would-be equilibrium dust temperature
$T_{\text{eq}}$ of the dust mixture, found by solving the equation
\begin{equation}
  \int_0^\infty 
  \kappa_\lambda^{\text{abs}}
  B_\lambda(T_{\text{eq}})\,\txd\lambda
  =
  \int_0^\infty 
  \kappa_\lambda^{\text{abs}}
  J_\lambda\,\txd\lambda
  \label{Teq}
\end{equation}
As a second parameter, we take the first-order moment
$\kappa_\lambda^{\text{abs}}J_\lambda$, i.e.\ the mean wavelength,
\begin{equation} 
  \bar{\lambda}
  =
  \frac{\int_0^\infty\kappa_\lambda^{\text{abs}}J_\lambda\,\lambda\,\txd\lambda}
  {\int_0^\infty \kappa_\lambda^{\text{abs}}J_\lambda\,\txd\lambda}
\end{equation}
In SKIRT we limit ourselves to two parameters, but in principle this
procedure can be extended to more parameters. 

With $T_{\text{eq}}$ and $\bar{\lambda}$ calculated in each dust cell,
we construct a 2D rectangular grid with $N_{T_{\text{eq}}} \times
N_{\bar{\lambda}}$ pixels in the $(T_{\text{eq}},\bar{\lambda})$
parameter space, based on the range of values encountered in the
present simulation. In every parameter space pixel we construct a
reference ISRF by averaging all ISFRs that correspond to those
particular values of $T_{\text{eq}}$ and $\bar{\lambda}$. We
experimented with different ways of averaging, including taking the
straight mean, the median or the mean using sigma-clipping, but found
no noticeable difference. The final step of the library construction is
to feed the reference ISRFs to the DustEM routine, and save each of
the resulting dust emissivity profiles in the library.

Once the library is constructed, finding the correct dust emissivity
for a given dust cell is straightforward, as each dust cell is already
connected to a certain pixel in the $(T_{\text{eq}},\bar\lambda)$
parameter space and hence a dust emissivity profile in the library.

\section{Implementation details}
\label{implementation.sec}

While the very first version of SKIRT were written in
Fortran~77, the code is now completely written in ANSI C++ and
currently contains some 30\,000 lines of code. It uses the
object-oriented nature of the C++ language extensively to support a
strong modularity. The use of inheritance and abstract classes renders
the inclusion of new components (such as new density distributions for
the stars or dust, or new dust mixtures) straightforward. The DustEM
code is written in Fortran~95 and has been slightly adapted to be
coupled to SKIRT. The entire SKIRT code is driven by a
graphic user interface written in PyQt. Batch jobs can be run using a
command line version with XML input files.

An important implementation aspect of SKIRT is the
parallelism. Parallelism can typically work on two domains: data
parallelism focuses on distributing the data across different parallel
computing nodes, with the principle aim of enabling simulations that
need more memory consumption than is available on a single node. Task
or control parallelism focuses on distributing execution processes
(threads) across different parallel computing nodes with the principle
aim of decreasing the run-time of a program. Ideally, both approaches
can be combined. Monte Carlo radiative transfer codes are easily
parallelized in a task parallelism approach: the different levels of
iterations can easily be split over different nodes. SKIRT uses
the OpenMP protocol to support task parallelism on shared-memory
machines. One of the advantages of OpenMP parallelism is the
spectacularly low coding cost: less than 100 lines of code (on a total
of 30\,000) have been added to SKIRT to convert it from a
serial into a parallel code. The main parallelism is situated in the
loop over wavelength, which implies that SKIRT runs both the
stellar and dust phases at different wavelengths simultaneously.

\begin{figure*}
\centering
\includegraphics[width=\textwidth]{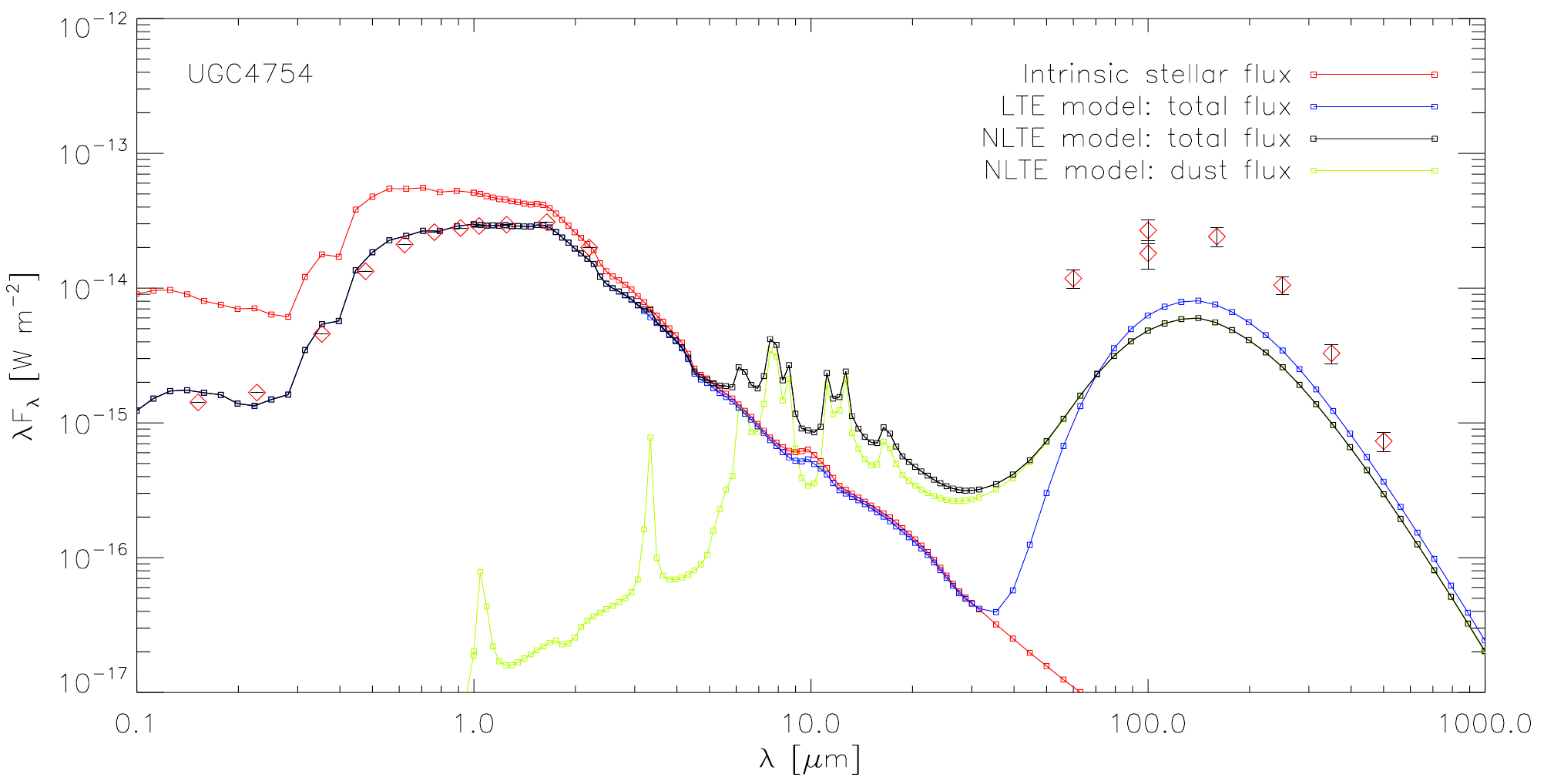}
\caption{SEDs of the SKIRT 2D models for UGC\,4754. The blue
  curve corresponds to a model assuming LTE dust re-emission, the
  black curve shows the SED of the model that includes NLTE dust
  re-emission. For the latter model, the yellow line shows the
  contribution of the thermal dust emission to the SED. The red curve
  represents the intrinsic flux of the model without dust extinction,
  and the data points are the observed total fluxes of UGC\,4754.}
\label{TotalSED.pdf}
\end{figure*}

One of the planned future developments of SKIRT is to look into
the possibilities of using the MPI interface for data parallelism, in
order to allow Monte Carlo simulations to be run on distributed memory
systems. Data parallelism is much harder to achieve for Monte Carlo
simulations than task parallelism. The main reason is the non-local
nature of the physical problem: each photon package in a simulation
typically requires data from the entire physical domain (read access
to calculate the optical depth and write access to update the
absorption rates). Contrary to e.g.\ grid-based hydrodynamic codes,
distributing the dust cells over different parts of memory would imply
an enormous overhead in communication between the different nodes. In
principle, data parallelism could be achieved by splitting the data in
the wavelength dimension, where different nodes contain different
parts of the absorption rate counters of the dust grid and different
parts of the detectors. However, in this approach a large amount of
data (such as the dust density grid) would need to be shared/copied
between the different nodes and communication overheads would be
significant in the dust emission phase. Future work will investigate
whether the benefits of distributed-memory parallelism can outweigh
the communication overheads and the significant additional coding
complexity.

\section{Tests and applications}
\label{applications.sec}

We have run extensive tests to check the accuracy of the SKIRT
code against other radiative transfer codes. The early versions of the
code were already tested against several other results, most
importantly the set of spiral galaxy models by
\citet{1994ApJ...432..114B} \citep[see
e.g.][]{2003MNRAS.343.1081B}. We have successfully tested the LTE
version of SKIRT against the 1D and 2D LTE circumstellar
benchmark problems of \citet{1997MNRAS.291..121I} and
\citet{2004A&A...417..793P}. SKIRT is also one of the codes
used in a new ongoing LTE benchmark effort focusing on a disc galaxy
environment (Baes et al.\ 2011, in preparation). The preliminary
results, based on the results from five independent radiative transfer
codes indicate excellent agreement, with relative differences in the
SEDs around the 1\% level or even below.

As a full NLTE radiative transfer benchmark is not (yet) available at
the moment, we have tested our NLTE radiative transfer code, and
particularly the library approach, using different models with
gradually increasing levels of complexity. In order to run simulations
in a realistic setting, we adopt the Sbc galaxy UGC\,4754 as a
template model. UGC\,4754 is a edge-on spiral galaxy, which has always
been a favorite class of galaxies for radiative transfer modellers, as
the dust is clearly visible both in absorption and emission
\citep[e.g.][]{1997A&A...325..135X, 1998A&A...331..894X,
  1999A&A...344..868X, 2000A&A...362..138P, 2011A&A...527A.109P,
  2001A&A...372..775M, 2005A&A...437..447D, 2007A&A...471..765B,
  2008A&A...490..461B}. This galaxy was one of the first large edge-on
galaxies to be observed with the Herschel Space Observatory as part of
the Herschel Astrophysical Terahertz Large Area Survey
\citep[H-ATLAS,][]{2010PASP..122..499E} science demonstration phase
observations. In \citet{2010A&A...518L..39B}, we fitted a radiative
model to the observed images of UGC\,4754 in the SDSS and UKIDSS
bands. We subsequently used the SKIRT code to predict the
galaxy's SED and images at FIR wavelengths. While the radiative
transfer model used in that paper was sufficient to serve its goal
(investigation of the dust energy balance), it suffered from two
significant limitations: the assumptions of LTE dust emission and of a
smooth interstellar medium. These two assumptions prevented us from
making a self-consistent model covering the entire spectral region
from the UV to mm wavelengths. Moreover, they also might introduce a
significant source of uncertainty, as it has been demonstrated by
several authors that the extinction properties of a clumpy
interstellar medium can be significantly different from those of a
smooth medium \citep[e.g.][]{1993MNRAS.264..145H,
  1996ApJ...463..681W,2000ApJ...528..799W, 1998A&A...340..103W,
  2000MNRAS.311..601B, 2001ApJ...548..150M, 2004ApJ...617.1022P,
  2005MNRAS.362..737D}.

In this section we will gradually refine our model for UGC\,4754 from
a smooth, 2D, LTE model to a fully 3D model that includes NLTE dust
emission and a clumpy structure of the dusty ISM. The main objectives
are to test the accuracy of our approach using a realistic setting,
and to demonstrate the ability of SKIRT to run realistic 3D NLTE
radiative transfer calculations. For a full investigation of the dust
energy balance in spiral galaxies, based on our refined SKIRT code and
multi-wavelength imaging data, we refer to future work.

\subsection{2D radiative transfer models}

The starting point for our models is the best fitting, smooth 2D model
from \citet{2010A&A...518L..39B}. The stellar distribution consists of
two components: a double-exponential stellar disc with a scale length
of 4.05~kpc and a scale height of 330~pc, and a flattened S\'ersic
bulge with a major axis effective radius of 800~pc, a S\'ersic
parameter of 0.9 and an intrinsic flattening of 0.6. Both components
have a similar intrinsic SED, corresponding to a population of 8 Gyr
old with an exponentially decaying star formation rate and an initial
burst duration of 0.15 Gyr. The total bolometric luminosity of the
system is $1.8\times10^{10}~ L_\odot$, of which the bulge contributes
8\%. The dust is also distributed in a double-exponential disc with a
scale length of 6.1~kpc and a scale height of 270~pc. The total amount
of dust is characterized by the $g$-band edge-on optical depth of
0.73, which corresponds to a total dust mass of
$1.0\times10^7~M_\odot$. Contrary to \citet{2010A&A...518L..39B} where
we used the \citet{2004ApJS..152..211Z} model to describe the dust
optical properties, we now use the \citet{2007ApJ...657..810D} dust
model, as this model is embedded in the DustEM library
\citep{2011A&A...525A.103C}.

Simulations are run on a wavelength grid with 181 grid points, with
101 grid points distributed logarithmically beween 0.05 and
5000~$\mu$m and 80 additional grid points distributed logarithmically
between 1 and 30~$\mu$m to capture the PAH peaks. For our 2D
simulations, we considered an axisymmetric grid with 51 grid points in
the radial direction and 51 grid points in the vertical direction,
resulting in a total number of $N_{\text{cells}} = 2500$ grid
cells. The grid points are chosen to have a power-law distribution,
with an extent of 30~kpc (2~kpc) for the radial (vertical)
distribution and a size ratio of 30 between the innermost and
outermost bins. In all SKIRT runs discussed here, we used
$10^6$ photon packages for each wavelength in both the stellar and the
dust emission phase.

Figure~{\ref{TotalSED.pdf}} shows the resulting SEDs of two different
SKIRT 2D simulations based on this model setup, as well as the
observed GALEX, SDSS, UKIDSS, IRAS and Herschel fluxes for
UGC\,4754. The blue curve shows the SED corresponding to a model
assuming LTE dust re-emission, where we took into account that
different grain types and sizes reach different equilibrium
temperatures. The black curve shows the SED of the model that includes
NLTE dust re-emission using the library approach discussed in
Section~{\ref{dustemissivity.sec}}. The red curve represents the flux
of the model without dust extinction, the yellow line corresponds to
the contribution of the dust to the SED in the NLTE case. It is
logical that there is no difference between the LTE and NLTE models in
the UV/optical/NIR part of the SED, as only the dust emissivity
differs between the models. There is, however, a significant
difference between the LTE and NLTE models in the MIR/FIR/submm
window: in the LTE models, all absorbed radiation is re-emitted as
modified blackbody emission in the FIR/submm region, whereas the NLTE
models emit part of the absorbed radiation in the MIR region.

\begin{figure}
\centering
\includegraphics[width=0.48\textwidth]{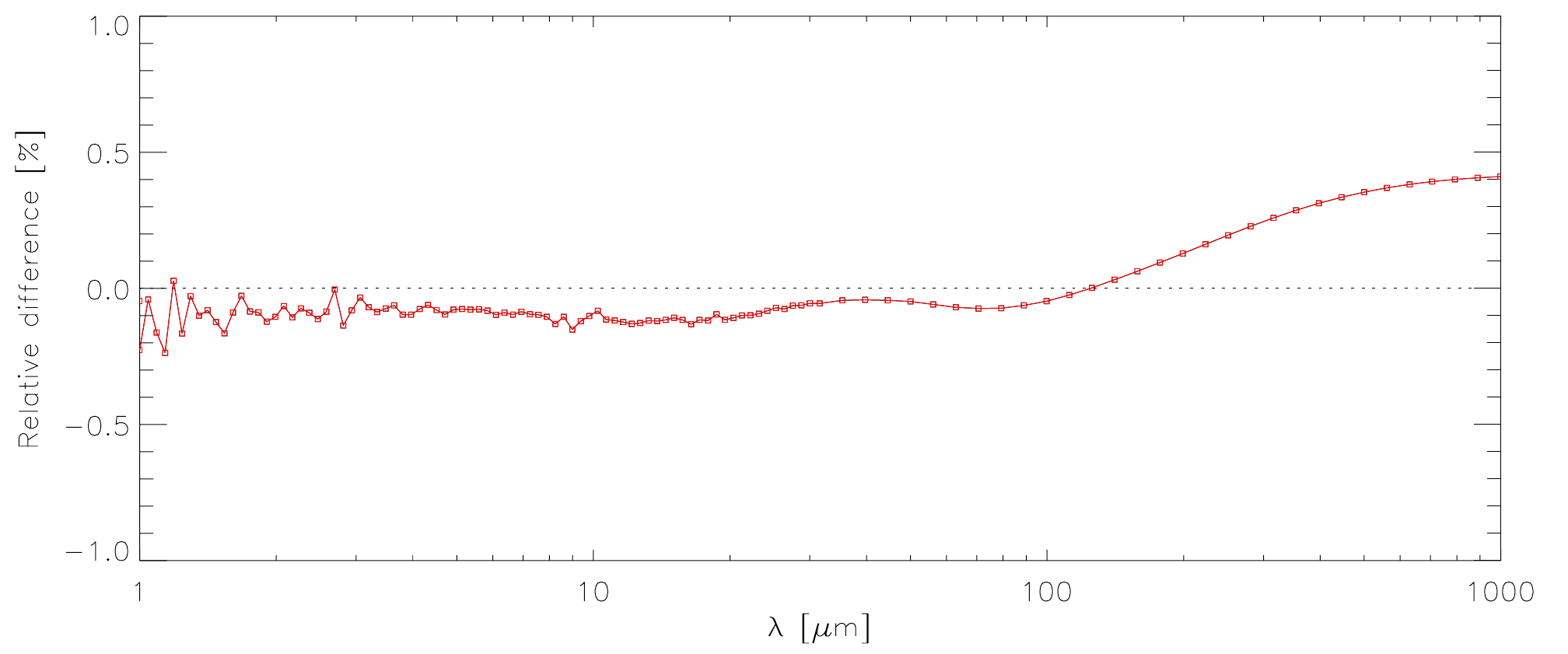}
\caption{The relative difference
  $(F_\lambda^{\text{bf}}-F_\lambda^{\text{lib}})/F_\lambda^{\text{lib}}$
  between the dust SEDs of 2D NLTE models for UGC\,4754 using the
  brute force (bf) approach and the library (lib) approach.}
\label{BruteForce.pdf}
\end{figure}

To test the accuracy of our library approach, we ran a second NLTE
simulation, where we did not use the library approach, but where we
calculated the dust emissivity in each cell by an explicit call to the
DustEM code. The relative difference between the SEDs of the
SKIRT NLTE models using the library approach and the brute
force approach is shown in Figure~{\ref{BruteForce.pdf}}. This figure
convincingly shows that the library approach is accurate: the relative
error is everywhere below 0.2\% and in the region where the dust
emission dominates even smaller. The difference in run time between
the two approaches is substantial, and this difference is only due to
the difference in the number of calls to the DustEM routine. In the
brute force approach, the number of calls is obviously equal to
$N_{\text{cells}}$ (or sometimes a bit less if there are empty dust
cells). The maximum number of calls to the DustEM routine in the
library approach is obviously $N_{T_{\text{eq}}} \times
N_{\bar{\lambda}}$, the number of pixels in the library parameter
space. As the intensity of the ISRF is the main parameter that drives
the shape of the dust emissivity spectrum \citep{2001ApJ...551..807D,
  2011A&A...525A.103C}, it is appropriate to choose
$N_{T_{\text{eq}}}$ somewhat larger than $N_{\bar{\lambda}}$; the
values we propose are $N_{T_{\text{eq}}} = 25$ and
$N_{\bar{\lambda}}=10$. This infers a maximum of 250 calls to the
DustEM routine. In most cases, however, not the entire
$(T_{\text{eq}},\bar{\lambda})$ parameter space is covered, such that
the number of calls is even further reduced. In the present
simulation, we needed to call the DustEM routine only 71 times in the
library approach, compared to 2500~times in the brute force
approach. Given that the average DustEM run time is about 7~s (on a
typical desktop computer) and that the overhead of the construction of
the library is negligible, this makes a substantial difference. It
should be noted that 2500 calls is still a very manageable number, but
the brute force approach becomes impossible when moving to 3D grids
with several million cells. For example, for a simulation with ten
million dust cells, we would need a DustEM computation time of more
than 2 years on a single core computer, compared to several minutes
using the library approach.

\begin{figure}
\centering
\includegraphics[width=0.48\textwidth]{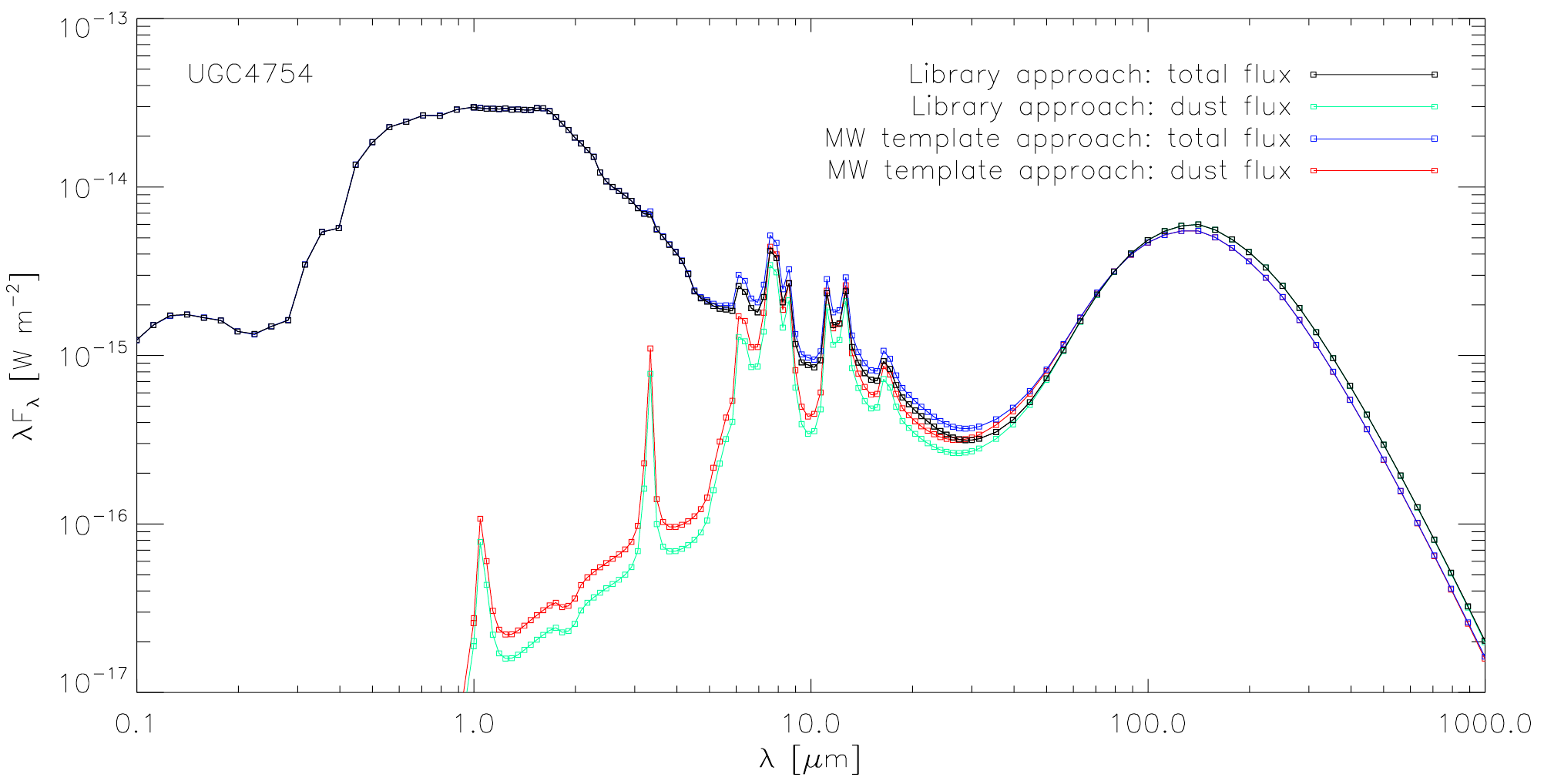}
\includegraphics[width=0.48\textwidth]{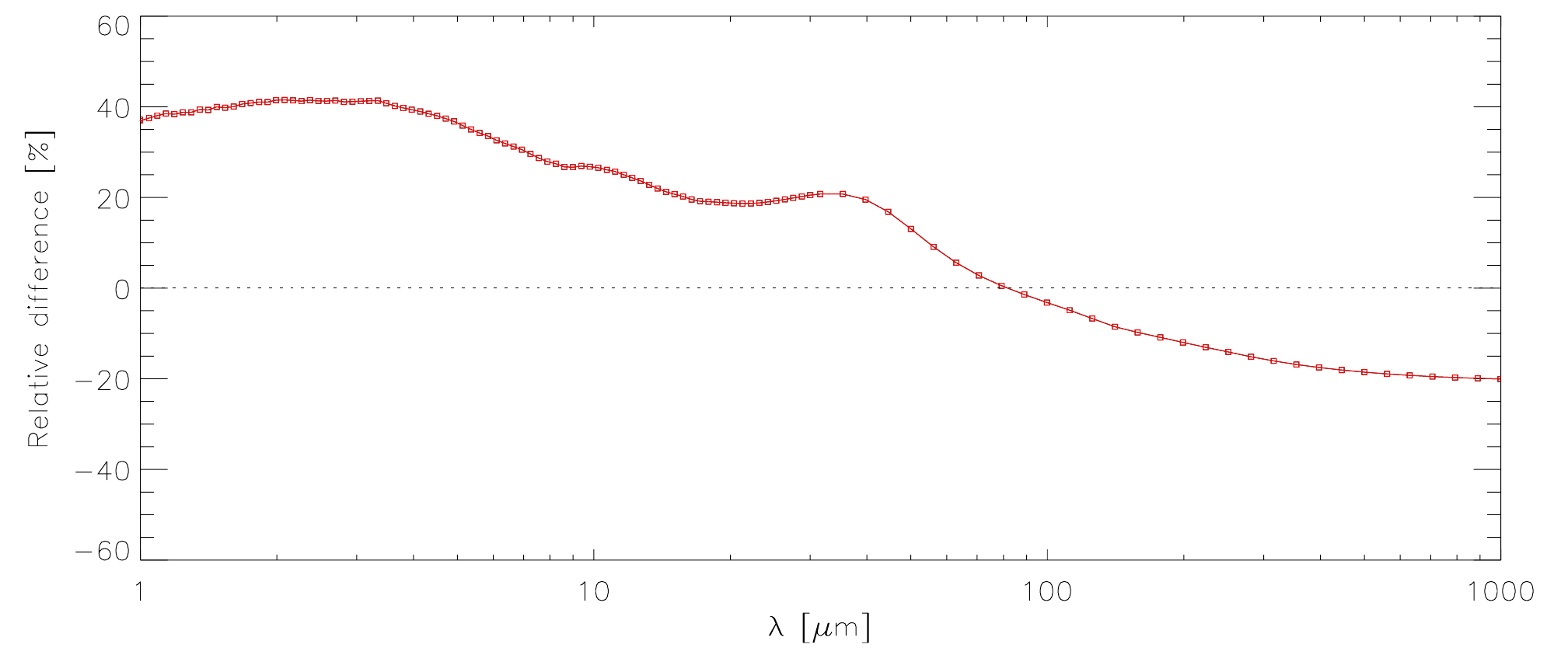}
\caption{Comparison of our library approach and the Milky Way ISRF
  template approach for 2D NLTE models for UGC\,4754. The top panel
  shows the total SEDs and the dust emission SEDs corresponding to
  both approaches, the top panel shows the relative difference
  $(F_\lambda^{\text{tem}}-F_\lambda^{\text{lib}})/F_\lambda^{\text{lib}}$
  between the dust SEDs using the MW ISRF template (tem) approach and
  our library (lib) approach.}
\label{MilkyWayLibrary.pdf}
\end{figure}

As a final test and a demonstration of the necessity of our library
approach, we ran our 2D simulation again, but now using a set of
precomputed template emissivity profiles. In a sense this mimics the
approach taken by e.g.\ \citet{2008ApJ...688.1118W}, although their
approach is somewhat different as they calculate the emission by big
grains using the LTE approximation and only use a template for the
small grains. The basic idea of our template approach is that the dust
emissivity from a cell depends only on a single parameter $U$, being
the integrated mean intensity of the ISRF in the considered cell
expressed in terms of the integrated mean intensity of the ISRF in the
Milky Way,
\begin{equation}
  U
  =
  \frac{\int_0^\infty J_\lambda\,\txd\lambda}
  {\int_0^\infty J_\lambda^{\text{MW}}\,\txd\lambda}
\end{equation}
For the ISRF of the MW, the standard parameterization of
\citet{1983A&A...128..212M} is adopted. We constructed a library of
501 dust emissivity profiles, corresponding to values of $U$
distributed logarithmically between $U_{\text{min}}=10^{-5}$ and
$U_{\text{max}}=10^5$. These dust emissivity profiles can be computed
once and for all (using the DustEM routine) and saved in a file. In
the dust emission phase, we simply calculate $U$ in every dust cell
and determine the dust emissivity profile using logarithmic
interpolation between the precomputed library profiles. This approach
definitely has a strong appeal: it is straightforward to implement and
very fast, as it requires only the calculation of $U$ and a simple
interpolation of precomputed values. In this sense it is more
attractive than our library approach, which requires the construction
of a library for every simulation. The disadvantage of the template
approach, however, is that only the strength and not the shape or
hardness of the ISRF is taken into account to calculate the dust
emissivity. As also argued by \citet{2010MNRAS.403...17J}, the shape
of the ISRF can have a significant importance, both because of the
differing cross-sections of grains of different sizes and because
high-energy photons will excite larger thermal fluctuations than
low-energy photons for a given value of $U$.

In Figure~{\ref{MilkyWayLibrary.pdf}} we show the comparison of the
fixed template approach and our dynamic library approach (dynamic in
the sense that the library is tailored to the specific
simulation). The top panel shows the total SED and the dust emission
SED for our 2D model for UGC\,4754 obtained using both approaches. At
first sight, the SEDs agree fairly well (definitely when plotted in
log-log scale). Looking at the bottom panel, where we plot the
relative difference of the dust emission SED corresponding to both
approaches, we do see a significant difference, with relative
deviations up to more than 40\%. Here we have to give an important
side note, in the sense that we believe that this 40\% difference is
actually a underestimate of the error one can make. The shape of the
intrinsic SED of the stellar population in our model is independent of
position; as a result, the variations in the shape of the ISRF in
different cells in the simulation are only the result of varying
levels of absorption and scattering. Since the stellar population
model we adopted for UGC\,4754 (an 8~Gyr old population with an
exponentially decaying star formation rate and an initial burst
duration of 150 Myr) is not very different from the average stellar
population in the Milky Way, this implies that the shape of the ISRF
in the different cells in our model will on average be quite close to
the shape of the Milky Way ISRF. As far as the comparison between our
library approach and the Milky Way template approach concerns, we must
hence conclude that UGC\,4754 simulations are not the strongest
test. For systems with ISRFs which deviate much more from the average
Milky Way ISRF, such as starburst galaxies or circumstellar discs
around young hot stars, we expect much larger differences than the
40\% we obtained here.

\subsection{3D clumpy radiative transfer models}

Having tested the accuracy of our library approach, we are ready to
run full-scale 3D NLTE radiative transfer models with
SKIRT. The first 3D model we consider has exactly the same
set-up as the 2D NLTE model, except that we now consider a uniform 3D
cartesian grid. In the $x$ and $y$ directions we consider 401 grid
points each and a maximum extent of 30~kpc, in the vertical direction
we use 61 grid points with a maximum extent of 2~kpc. This results in
$N_{\text{cells}} = 9.6\times10^6$ grid cells, each with a dimension
of 150~pc in the $x$ and $y$ directions and 66.7~pc in the vertical
direction. Note that we need to store, at each dust grid cell, the
entire ISRF $J_\lambda$ at each of the wavelength grid points, which
basically turns our grid into a 4-dimensional grid structure with
$1.73\times10^9$~grid cells. The memory required to run such a
large-scale SKIRT radiative transfer simulation is about 23~GB.

\begin{figure}
\centering
\includegraphics[width=0.48\textwidth]{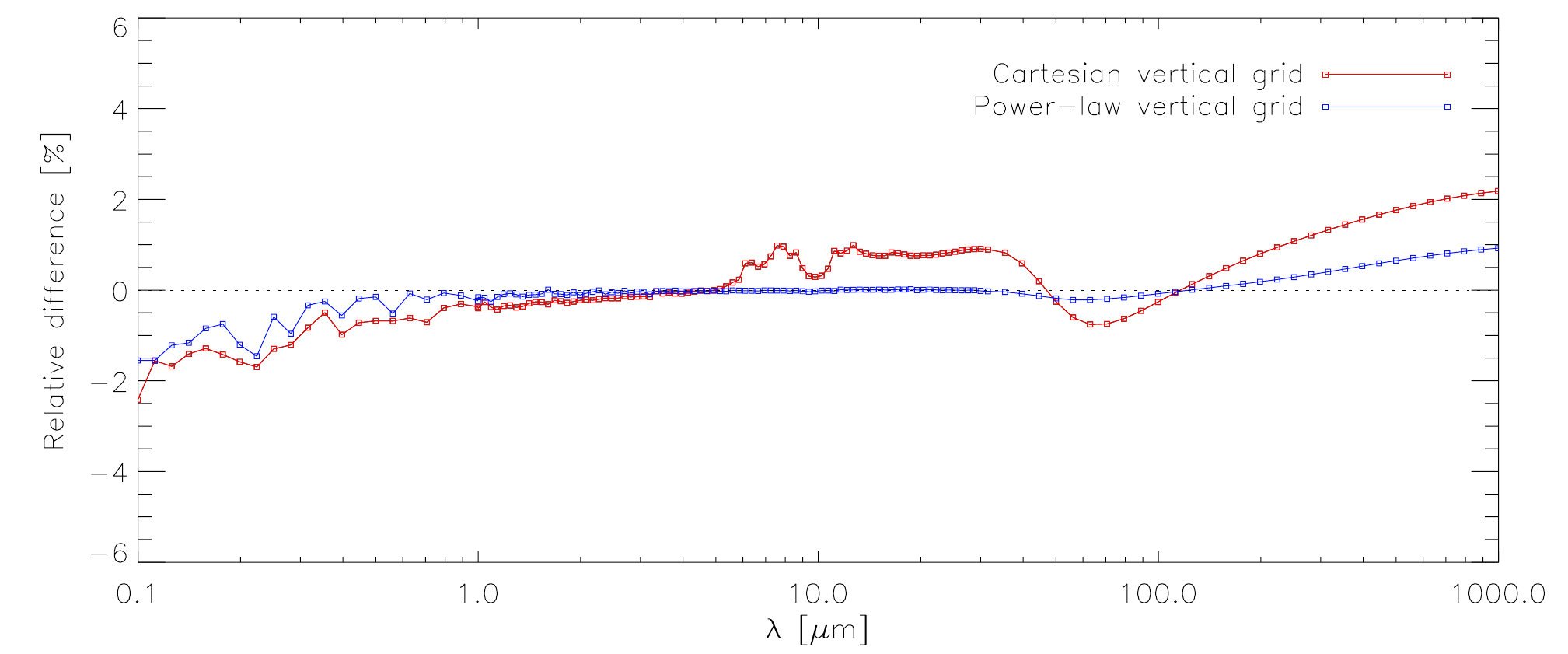}
\caption{The relative difference
  $(F_\lambda^{\text{3D}}-F_\lambda^{\text{2D}})/F_\lambda^{\text{2D}}$
  between the SEDs of 2D and 3D NLTE models for UGC\,4754. The red
  curve corresponds to a 3D model with a uniform cartesian dust grid
  in all three dimensions, the blue curve corresponds to a 3D model
  with a power-law grid in the vertical direction, similar to the
  vertical grid structure of the 2D model.}
\label{Compare2D3D.pdf}
\end{figure}

Figure~{\ref{Compare2D3D.pdf}} compares the SED of the smooth 2D and
3D models for UGC\,4754. The relative differences are below 2\% in the
entire UV-mm domain. The existing small differences are mainly due to
the discretization of the grid in the vertical direction: for the 2D
model we used a vertical grid with a power-law distribution, with
smaller bins close to the equatorial plane, where the dust density has
the strongest gradients. The innermost grid cell has a height of only
8.75~pc, compared to the 66.7~pc in the case of the 3D grid. The
result is that the discretization of the dust density on the 3D grid
is much coarser. To demonstrate that this vertical grid distribution
is the origin of this $<2\%$ difference, we ran another 3D simulation
where we now apply the same power-law distribution for the vertical
grid cells as we did for the 2D model. The relative differences
between the SED of this model and the SED of the 2D simulation are
also indicated in Figure~{\ref{Compare2D3D.pdf}}.

The reason why we considered a uniform cartesian dust grid is because
such a grid forms the basis for a fully 3D model with a clumpy,
two-phase dust distribution. To generate such a model, we followed the
strategy outlined by \citet{1996ApJ...463..681W}. The dusty
interstellar medium consists of two phases, a smooth inter-clump
component and a clumpy component, and is characterized in terms of two
parameters, namely the volume filling factor {\textit{ff}} of the
dense clumps and the density contrast $C$ between the clump and
inter-clump medium. The practical construction of the two-phase medium
consists of randomly assigning a status (clump or inter-clump) to each
dust cell in the dust medium. Typical values for the parameters $C$
and {\textit{ff}} vary widely in the literature. We use the values
$C=100$ and ${\textit{ff}}=0.1$ in this work, which are within the
range of typical parameters used in other studies
\citep[e.g.][]{1998AJ....115.1438K, 2000ApJ...528..799W,
  2000MNRAS.311..601B, 2001ApJ...548..150M, 2004ApJ...617.1022P,
  2005MNRAS.362..737D}.

\begin{figure*}
\centering
\includegraphics[width=0.8\textwidth]{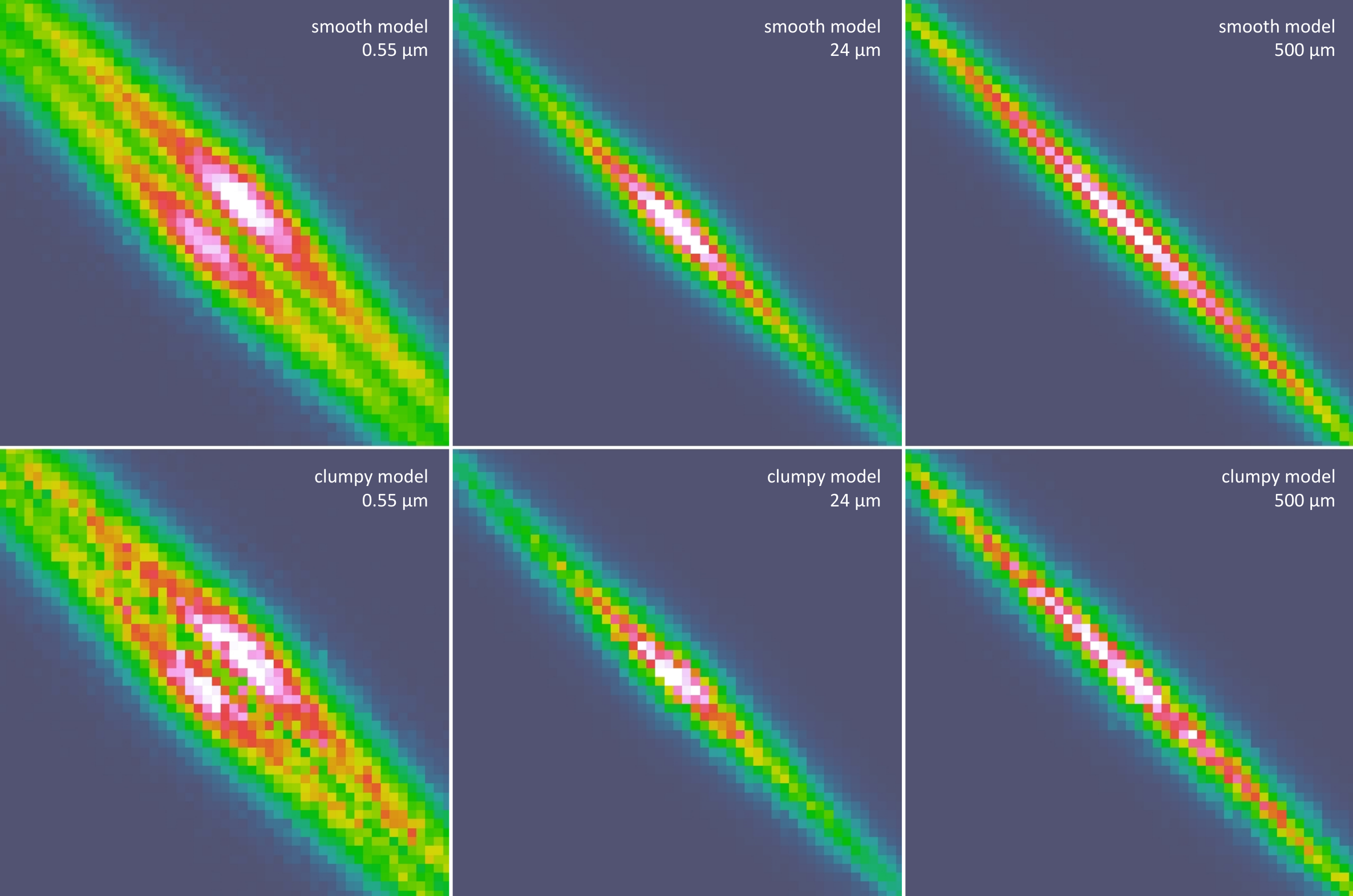}
\caption{Simulated images corresponding to the smooth (top row) and
  clumpy (bottom row) 3D models for UGC\,4754. The three panels
  correspond to three different regimes: stellar emission and dust
  extinction (0.55~$\mu$m, left), NLTE emission by small grains
  (24~$\mu$m, middle) and LTE emission by big grains (500~$\mu$m,
  right).}
\label{ClumpySmooth.pdf}
\end{figure*}

\begin{figure}
\centering
\includegraphics[width=0.48\textwidth]{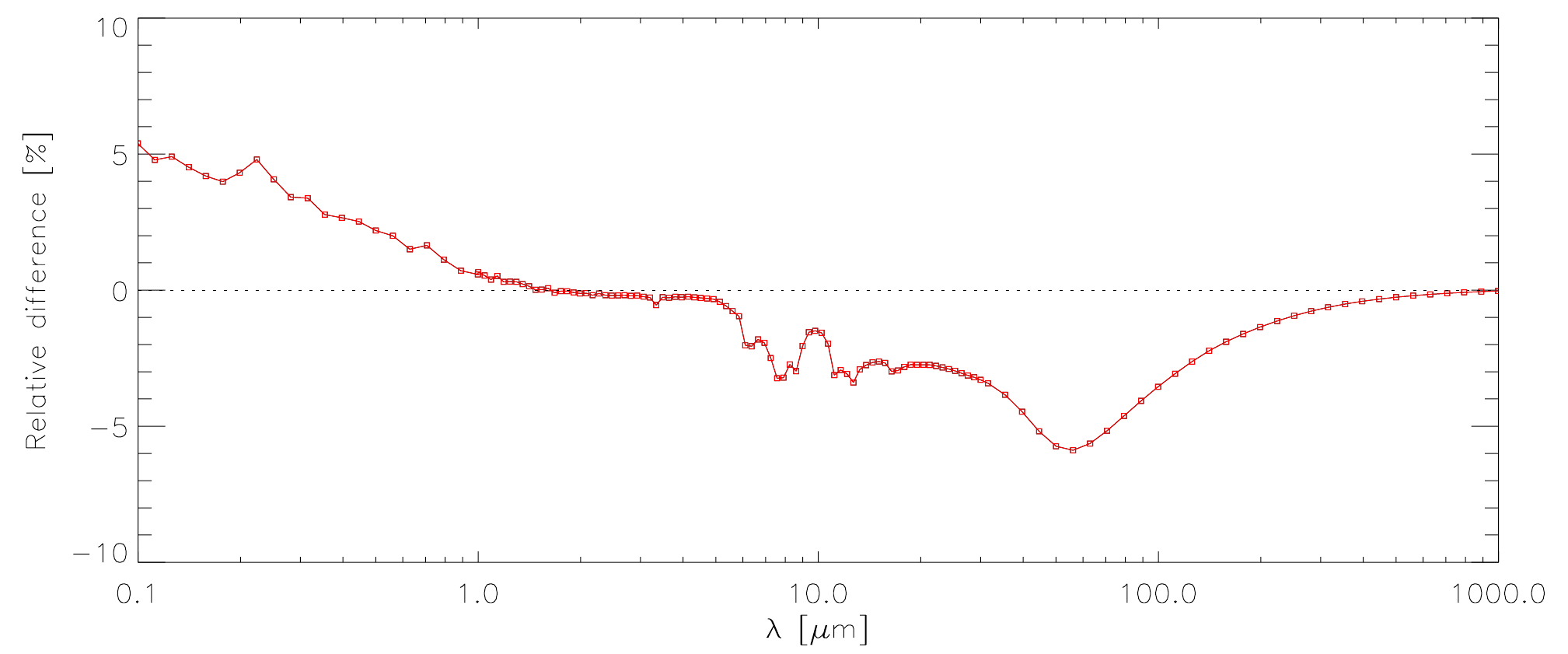}
\caption{The relative difference
  $(F_\lambda^{\text{clumpy}}-F_\lambda^{\text{smooth}})/F_\lambda^{\text{smooth}}$
  between the SEDs of clumpy and smooth 3D NLTE models for UGC\,4754.}
\label{CompareClumpySmooth.pdf}
\end{figure}

In Figure~{\ref{ClumpySmooth.pdf}}, we show simulated images
corresponding to the clumpy and smooth 3D models for the inner region
of UGC\,4754 at three different wavelengths (0.55, 24 and
500~$\mu$m). It is important to note that the pixel-to-pixel
variations in the images on the bottom row are not due to Poisson
noise in the Monte Carlo routine, but represent real intensity
variations due to the clumpy nature of the dust in the
models. Figure~{\ref{CompareClumpySmooth.pdf}} compares the total SED
of the clumpy and smooth models. The relative differences are below
6\% in the entire UV-mm domain. At UV wavelengths, the clumpy model is
slightly brighter than the smooth model, or put differently, slightly
less UV radiation is absorbed by the dust. This is in agreement with
conclusions found by other authors: for a fixed amount of dust, a
clumpy dust medium absorbs radiation less efficiently than a smooth
dust medium \citep[e.g.][]{2000MNRAS.311..601B,
  2004ApJ...617.1022P}. When moving to longer wavelengths, the
difference between the smooth and clumpy models decreases, as the
fraction of absorbed versus unattenuated radiation decreases with
increasing wavelength. At NIR wavelengths, there is virtually no
difference anymore between the SEDs of the smooth and clumpy
models. Moving to MIR and FIR wavelengths, we find that the clumpy
model emits significantly less, particularly at wavelengths up to
about 100~$\mu$m. This is no surprise, as the clumpy model was less
efficient in absorbing UV and optical radiation. The result is that,
for the same total mass, the dust in a clumpy model is on average both
cooler and less luminous than the dust in a smooth model.

\section{Conclusions}
\label{conclusions.sec}

We have presented an updated version of the 3D Monte Carlo radiative
transfer code SKIRT. The code uses various advanced
optimization techniques, both well-known and novel ones, that make the
Monte Carlo process orders of magnitude more efficient than the most
basic Monte Carlo technique. These techniques include an optimized
combination of eternal forced scattering and continuous absorption, a
multi-gaussian expansion technique and an efficient foam generator to
generate random positions from the stellar density, and the use of
peeling-off and smart detectors to create high signal-to-noise images
and SEDs.

The main novelty of the new SKIRT code is the possibility to
calculate the dust temperature distribution and the associated
infrared and submm emission with a full incorporation of the emission
of transiently heated grains and PAH molecules. To achieve this, we
have chosen to link the SKIRT code to DustEM
\citep{2011A&A...525A.103C}, a publicly available, state-of-the-art
numerical tool designed to calculate the NLTE emission of arbitrary
mixtures of dust grains. The advantages of this approach is that no
LTE approximation is made, even for large grains, and that new physics
(such as spinning dust emission or a temperature-dependent dust
emissivity) can readily be included. We have implemented a library
approach to limit the computational cost of the NLTE dust emission
calculations inherent in DustEM. Our approach is inspired by the work
by \citet{2003A&A...397..201J}, but uses a slightly different approach
that makes maximum use of all information in the simulation to
calculate the dust emissivity and avoids the need for additional
low-resolution simulations.

We have tested the accuracy of the SKIRT code, in particular of
our NLTE library approach, through a set of simulations based on the
edge-on spiral galaxy UGC 4754, previously modelled by
\citet{2010A&A...518L..39B}. The models we ran were gradually refined
from a smooth, 2D, LTE model to a fully 3D model that includes NLTE
dust emission and a clumpy structure of the dusty ISM. 

Using 2D models, we demonstrated the accuracy of our library approach:
the relative differences in the SED between a model that uses the
library approach and a model that uses brute force to calculate the
dust emission are less than 0.2\% at all wavelengths. Even for this 2D
model with only 2500 dust cells, the difference in run time between
both approaches are substantial; for 3D grids with several million
dust cells the brute force approach becomes impossible. We have also
explored the possibility to use a fixed set of precomputed dust
emission templates instead of a dynamic library as the one we have
chosen. While a template approach has the advantage that it is easier
to implement and faster to run, we have demonstrated that it leads to
significant deviations due to the fact that it does not take into
account the shape of the interstellar radiation field. This highlights
the need for a more advanced approach such as the library approach we
propose.

We have subsequently applied the SKIRT code to calculate
full-scale 3D NLTE models for UGC 4754. We found small differences ($<
2$\%) between 2D and 3D smooth models that are mainly due to
differences in the vertical discretization of the internal
grid. Finally, we have compared 3D models with a smooth and a clumpy
interstellar dust medium. We confirm the result found by other authors
that, for a fixed amount of interstellar dust, a clumpy dust medium
absorbs radiation less efficiently than a smooth dust medium. As a
direct consequence, the dust in clumpy models is on average both
cooler and less luminous, and the observed infrared emission of clumpy
models is less than the emission at these wavelengths of smooth models
with the same dust mass.

Our simulations demonstrate that, given the appropriate use of
optimization techniques, it is possible to efficiently and accurately
perform Monte Carlo radiative transfer simulations of arbitrary 3D
structures of several million dust cells, including a full calculation
of the NLTE emission by arbitrary dust mixtures. This significantly
increases the number of applications where detailed radiative transfer
modeling can be used. For example, we have started an investigation
of the energy balance crisis in a set of edge-on spiral galaxies: our
intention is to fit detailed radiative transfer models to
UV/optical/NIR images for a set of edge-on spiral galaxies, predict
the resulting MIR/FIR/submm emission and compare these predictions
with the available long wavelength data. Many other applications
(AGNs, circumstellar discs, merging galaxies,\ldots) are possible, and
the authors welcome all projects that can make use of SKIRT.

\acknowledgements{M. B., J.~V.\ and E.~V.~P.\ acknowledge the support
  of the Fund for Scientific Research Flanders (FWO-Vlaanderen) that
  made the development of the SKIRT code possible. M.~B., I.~D.~L.\
  and J.~F.\ thank the Belgian Federal Science Policy Office (BELSPO)
  for the financial support. M. S.\ acknowledges support of the
  Ministry of Education and Science of the Republic of Serbia throught
  the projects ``Astrophysical Spectroscopy of Extragalactic Objects''
  (176001) and ``Gravitation and the Large Scale Structure of the
  Universe'' (176003) and support of the European Commission (Erasmus
  Mundus Action 2 partnership between the European Union and the
  Western Balkans, \mbox{http://www.basileus.ugent.be}) during his
  mobility period at Ghent University.}

\end{document}